# HOW SUPERNOVAE BECAME THE BASIS OF OBSERVATIONAL COSMOLOGY

**Maria Victorovna Pruzhinskaya**
*Laboratoire de Physique Corpusculaire, Université Clermont Auvergne, Université Blaise Pascal, CNRS/IN2P3, Clermont-Ferrand, France; and Sternberg Astronomical Institute of Lomonosov Moscow State University, 119991, Moscow, Universitetsky prospect 13, Russia.*
Email: pruzhinskaya@gmail.com

**and**

**Sergey Mikhailovich Lisakov**
*Laboratoire Lagrange, UMR7293, Université Nice Sophia-Antipolis, Observatoire de la Côte d'Azur, Boulevard de l'Observatoire, CS 34229, Nice, France.*
Email: lisakov57@gmail.com

**Abstract:** This paper is dedicated to the discovery of one of the most important relationships in supernova cosmology—the relation between the peak luminosity of Type Ia supernovae and their luminosity decline rate after maximum light. The history of this relationship is quite long and interesting. The relationship was independently discovered by the American statistician and astronomer Bert Woodard Rust and the Soviet astronomer Yury Pavlovich Pskovskii in the 1970s. Using a limited sample of Type I supernovae they were able to show that the brighter the supernova is, the slower its luminosity declines after maximum. Only with the appearance of CCD cameras could Mark Phillips re-inspect this relationship on a new level of accuracy using a better sample of supernovae. His investigations confirmed the idea proposed earlier by Rust and Pskovskii.

## 1 INTRODUCTION

In 1998–1999 astronomers discovered the accelerating expansion of the Universe through the observations of very far standard candles (for a review see Lipunov and Chernin, 2012). In astronomy Type Ia supernovae (SNe) are used as standard candles. First, they are bright enough to be visible from cosmological distances. And second, their luminosity at maximum light is approximately the same. This last property derives from the fact that the Type Ia supernova (SN) phenomenon is an explosion of a carbon-oxygen white dwarf—an object with nearly a solar mass but a hundred times smaller in diameter. In this case we can calculate the brightness of a supernova at any distance within the framework of any cosmological model.

Thus, comparing the observations with theoretical predictions one determines the cosmological parameters of the Universe. Such measurements became possible with the development of observational tools and the emergence of major supernovae surveys (Dong et al., 2016; Gal-Yam et al., 2013; Lipunov et al., 2010; Rau et al., 2009).

From a theoretical point of view, the accelerating expansion of the Universe can be explained in Einstein's equations of General Relativity by the introduction of a cosmological constant, or more generally by an unknown form of energy with a negative pressure, so-called 'dark energy' (Einstein, 1917; 1997; McVittie, 1965).

However, from the moment that Albert Einstein (1879–1955; Whittaker, 1955) introduced into the equations of the General Theory of Relativity a cosmological constant until the discovery of the accelerating expansion of the Universe, nearly 80 years would pass.

## 2 A BRIEF ACCOUNT OF SUPERNOVAE

A hundred years ago our concept of the Universe was remarkably different from the modern view. Estimations of the size of our galaxy ranged from $5 \times 10^3$ to $3 \times 10^5$ l.y.; the location of our Solar System within our galaxy remained an open question; and there was no agreement about whether spiral nebulae belong to our galaxy or not. These topics were discussed during 'The Great Debate' between Harlow Shapley (1885–1972; Trimble and Smith, 2014) and Heber Doust Curtis (1872–1942; Lindner and Marché, 2014) in 1920 (Shapley and Curtis, 1921). Objects which today are known as novae and supernovae were prominent characters in this dispute. However, at that time all new star-like bright objects that gradually faded over a period of several months were referred to as 'novae'.

The tradition to assign this name (novae) to such objects was established back in the sixteenth century by Tycho Ottesen Brahe (1546–1601; Moesgaard, 2014) in his work titled "Concerning the new and previously unseen star" (our English translation of the Latin original) about the famous object SN 1572, a supernova in our galaxy that was visible to the naked eye. At that time, of course, it was not called a 'supernova'.

As van den Bergh (1988) reminds us, prior to 1917 only two novae had been observed in spiral nebulae, S Andromedae in 1885 and Z Centauri in 1895, whereas dozens of novae had been recorded in our own galaxy.[1] The discovery of a new star in NGC 6946 by George Willis Ritchey (1864–1945; Cameron, 2014) in 1917 motivated many observatories to start a search for such objects in archival data. As a result, Shapley (1917) published a list of eleven 'temporary stars' in spiral nebulae—today we know that three of these were actually novae, while the remaining eight were supernovae. Curtis (1917) adopted mean values $<m_{max}> \approx 5$ for galactic novae and $<m_{max}> \approx 15$ for novae in spiral nebulae (van den Bergh, 1988). Therefore, novae in spiral nebulae were on average ~10 magnitudes fainter than other novae. This could be explained in one of two different ways: either novae in spiral nebulae were generally $10^4$ times fainter, or novae in spiral nebulae were on average 100 times further away than other novae. Shapley doubted Curtis' idea that spiral nebulae were galaxies comparable in size to our own galaxy.[2] If one adopts this 'comparable galaxy theory', then distances to the spiral nebulae, estimated through their angular diameters, would be immense. Shapley noted among the points of agreement between himself and Curtis:

> If our galaxy approaches the larger order of dimensions, a serious difficulty at once arises for the theory that spirals are galaxies of stars comparable in size with our own: it would be necessary to ascribe impossibly great magnitudes to the new stars that have appeared in the spiral nebulae. (Shapley and Curtis, 1921: 180).

Curtis suggested that at most, spiral nebulae were ~10 times smaller than our galaxy. Thus, with this greatly-reduced scale of the Universe it was easier to accept the still impressively-great magnitudes of novae in spiral nebulae.

A few years after the Great Debate, a relationship between recessional velocity and distance to spiral nebulae was established. In 1927, Georges Henri-Joseph Edouard Lemaître (1894–1966; Kragh, 2014) published a work titled "A homogeneous Universe of constant mass and growing radius accounting for the radial velocity of extragalactic nebulae" (our English translation of the French original) in the *Annals of the Scientific Society of Brussels* (Lemaître, 1927). In this work he found dynamic solutions to Einstein's General Theory of Relativity equations. These solutions indicated that there was a linear relationship between the recessional velocity of a galaxy and its distance. Lemaître was the first to estimate the value of this parameter, linking the recessional velocity ($v$) and galaxy distance ($D$). Today this parameter is known as $H_0$ where

$$v = H_0 D \qquad (1)$$

Using redshift measurements of Vesto Melvin Slipher (1875–1969; Giclas, 2014) from Gustav Strömberg (1882–1962; Hockey and MacPherson, 2014) published in 1925 and distance estimations from Hubble (1926) for 42 galaxies, Lemaître obtained a value of 625 km/s/ Mpc. In a cited work, Hubble used a statistical expression for distance ($D$) in parsecs:

$$log(D) = 4.04 + 0.2m \qquad (2)$$

where $m$ is the total apparent magnitude. Lemaître noted, however, that contemporary uncertainties in distances to galaxies were too high to show this linear relationship between velocity and distance. His value of $H_0$ is quite different from the modern $H_0 = 67.80 \pm 0.77$ km/s/Mpc (Planck Collaboration, 2014), but close to the value of 500 km/s/Mpc obtained by Hubble two years later (Hubble, 1929), where he used Cepheid variables to find distances to the galaxies. The main reason of their significant errors was incorrectly-determined distances to galaxies. To determine distances, Hubble used the period-luminosity relation for Cepheid variables in nearby galaxies and the brightest resolved stars as the most luminous individual locators for distant galaxies. In 1952 it was shown that there were two types of Cepheid variables (Baade, 1952; Thackeray, 1952), and Allan Rex Sandage (1926–2010; Lynden-Bell and Schweizer, 2012) summarized problems associated with both approaches (Sandage, 1958).

Once the distances to the hosts of novae had been determined, it became possible to estimate their absolute magnitudes. It turned out that some novae were orders of magnitude brighter than others, and astronomers realized that novae should be divided into two subclasses. This division occurred in 1934, when Fritz Zwicky (1898–1974; Knill 2014; Figure 1) and Walter Baade (1893–1960; Florence, 2014; Figure 2) suggested the term 'super-novae' for 'exceptionally bright novae'.[3] It then took a couple of years for the hyphen to disappear. In 1938 Baade noticed that supernovae were a more homogeneous class of objects than novae. He found the mean absolute magnitude at maximum light for 18 supernovae[4] to be $-14.3^m$, with a dispersion of ~$1.1^m$. Therefore, supernovae were considered as good distance indicators in the Universe (Baade, 1938; Wilson, 1939; Zwicky, 1939). In 1941 Rudolph Leo Bernhard Minkowski (1895–1976; Durham, 2014)[5] obtained and analyzed the first spectra of supernovae, dividing them into two main types (Minkowski, 1941). To Type I he attributed supernovae that had no hydrogen lines in their spectra, where the

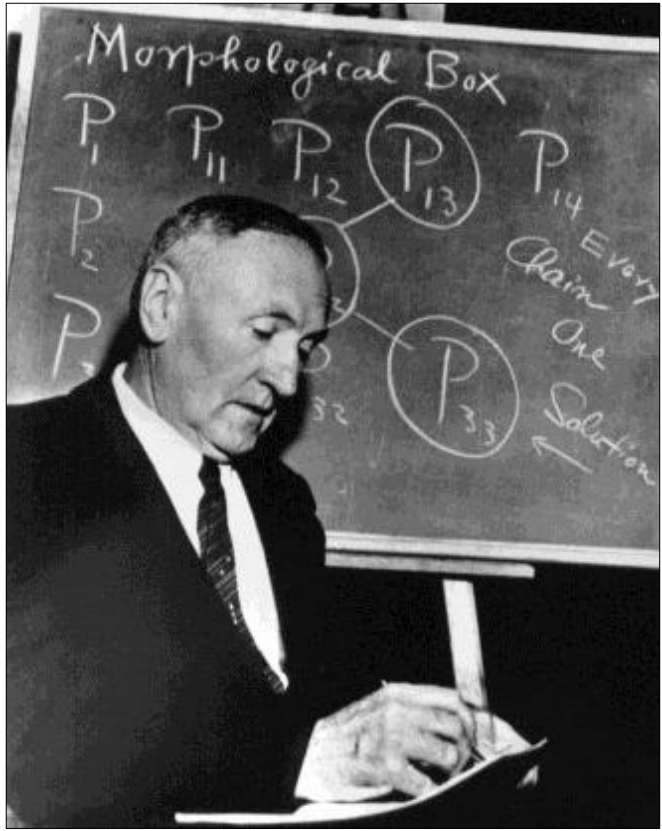
Figure 1: Fritz Zwicky (https://www.wikipedia.org)

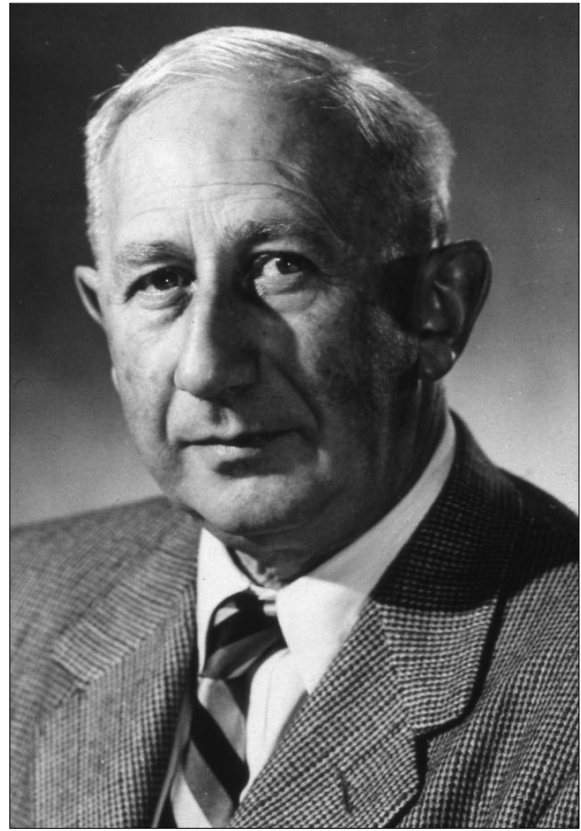
Figure 2: Walter Baade (https://www.wikipedia.org)

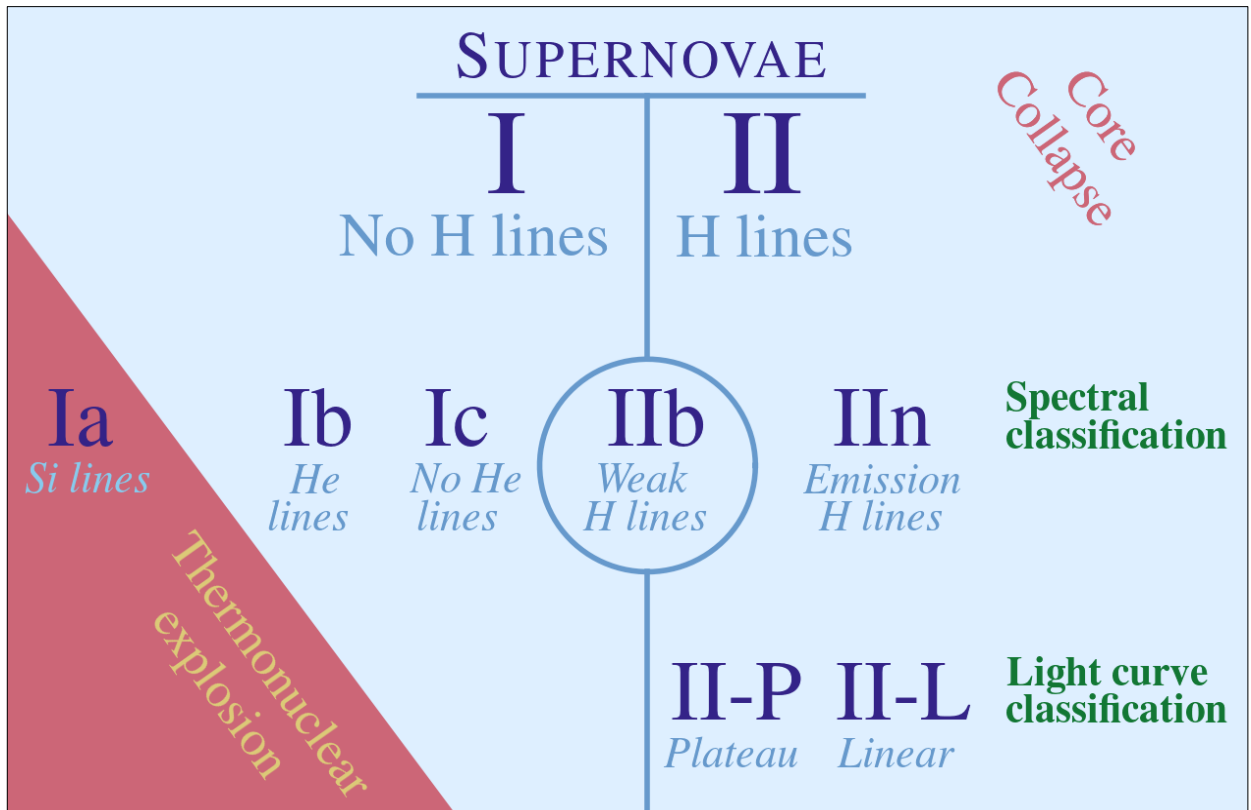
Figure 3: Supernova classification (drawn by authors)

entire spectrum consisted of broad maxima and minima that were not possible to explain.[6] Type II supernovae, on the contrary, showed the presence of hydrogen in their spectra. Over time, a more detailed classification appeared (see Figure 3). Type I supernovae were divided into three SN subtypes: Ia, Ib and Ic. It was found that the SN Ia phenomenon arises from the thermonuclear explosion of a white dwarf, and Type II SNe and SNe Ib/c from the core collapse of a massive star at the final stage of its evolution. Thus, SN Ia explosions differ in their origin from other supernovae.

As mentioned previously, the SN Ia phenomenon is an explosion of a carbon-oxygen white dwarf of nearly 1 $M_\odot$ but with a diameter hundred times smaller than the Sun. A thermonuclear explosion occurs when the mass of the white dwarf exceeds the Chandrasekhar Limit either by matter accretion from a companion star or by merging with another white dwarf. SNe Ia are highlighted by the presence in their spectra of absorption lines of singly-ionized silicon. During the explosion, ~0.5 $M_\odot$ of $^{56}$Ni is produced. The radioactive decay $^{56}$Ni $\rightarrow$ $^{56}$Co $\rightarrow$ $^{56}$Fe powers the maximum and post-maximum light curve of SNe Ia (Colgate and McKee 1969; Hoyle and Fowler, 1960).

To explain the spectra of core-collapse supernovae it is important to know which part of the envelope of the star is lost before the core collapse. If the stellar wind by which the star loses matter is not intense, the collapse occurs at the stage of the red supergiant. The radius of these stars can be several hundred times greater than the solar radius, and their extremely tenuous envelopes contain large amounts of hydrogen. Therefore, red supergiants are the progenitors of Type II SNe, in the spectra of which hydrogen lines are the most prominent.

More massive stars lose mass more efficiently by the stellar wind and end their lives losing all or part of the hydrogen envelope. Although having the same energy source, such core-collapse supernovae do not show hydrogen in their spectra (SNe Ib), or only show it in small quantities (SNe IIb). An even more effective stellar wind can 'blow out' not only the hydrogen but also the helium envelope. As a result, SNe Ic explode. Phenomenon of SNe Ib/c also may arise from an explosion in a binary system where the envelope is lost due to interaction with a companion star.

It was only when SNe were divided into Types and Subtypes that the most homogeneous SNe Ia could be identified and used as cosmological distance indicators.

Studies of supernovae led to one of the greatest discoveries in observational cosmology: the accelerating expansion of the Universe. In 1998–1999, two international teams of astronomers, one led by Brian Schmidt and Adam Riess and the other by Saul Perlmutter, reported that the cosmological expansion was accelerating.

The 'Supernova Cosmology Project' started in 1988 under the leadership of Saul Perlmutter, with the aim of determining the cosmological parameters of the Universe using the relationship of 'distance modulus–redshift' for distant SNe Ia. The first results, obtained for seven supernovae at redshift of $z$ ~0.4, gave a zero value for the cosmological constant. However, a more detailed analysis, including 42 cosmological supernovae with redshifts from 0.18 to 0.83, showed that in the case of a flat Universe ($\Omega_M + \Omega_\Lambda = 1$) the density of matter is $\Omega_M = 0.28$ +0.09/–0.08 (1-$\sigma$ statistical error) +0.05/–0.04 (systematic error). The probability that the dark energy density is not equal to zero was 99.8% (Perlmutter et al., 1999).

Brian Schmidt's competing project, involving the 'High-Z Supernova Search Team', was launched in 1995. Their first attempts to detect the accelerating expansion also were unsuccessful due to large measurement errors. Only in 1998, using an expanded sample of 16 distant supernovae, were they able to show that in the case of a flat Universe $\Omega_M = 0.28 \pm 0.1$ (Riess et al., 1998).

It is noteworthy that most of the distant supernovae in both projects were discovered at the Cerro Tololo Inter-American Observatory with the 4-meter Blanco Telescope and nearby supernovae with $z$ <0.1 were taken from the Calan/Tololo supernovae survey. The spectra of distant supernovae were obtained with the Keck telescopes by Alex Filippenko who was a member of both teams. In 2011, for their discovery of the accelerating expansion of the Universe through observations of distant supernovae Saul Perlmutter, Brian Schmidt and Adam Riess were awarded the Nobel Prize in Physics (Figure 4).

However, this important discovery would not have been possible without another discovery: that there was a relationship between the peak luminosity of an SN Ia and the shape of its light curve.

## 3 THE 'STANDARDIZATION' OF A CANDLE

The SNe Ia light curves in most cases are very similar to each other (see Figure 5). Approximately at 15 days, the luminosity of an SN Ia reaches a maximum, which then lasts for a few days. At maximum light SNe Ia have on average an absolute magnitude in the *B*-band of the order of –19.5$^m$. At that moment, the luminosity of the star is comparable to the luminosity of the entire host galaxy! After reaching maximum light the luminosity of an SN Ia declines rapidly, by ~3$^m$ in 25–30 days, following an almost linear increase of apparent magnitude, which corresponds to an exponential dimming of the luminosity (Tsvetkov et al., 2009). Nearby SNe Ia (e.g. SN 2011fe, SN 2014J) can be observed for about a year.

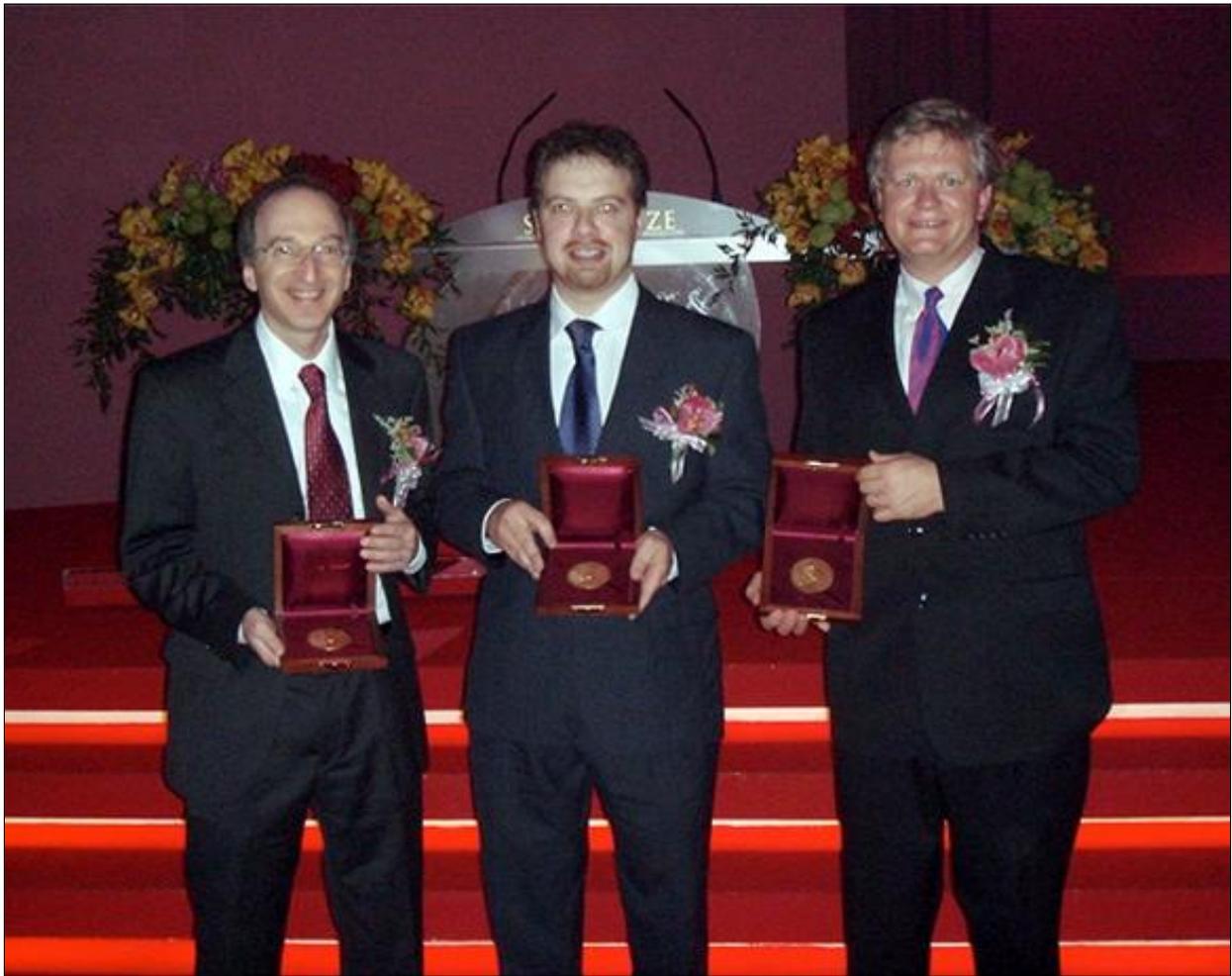

Figure 4: The Nobel laureates in Physics in 2011, Saul Perlmutter, Adam Riess and Brian Schmidt (from Wikipedia, Licensing: this work has been released into the public domain by its author, Ariess at English Wikipedia; this applies worldwide).

Despite the apparent similarities, the light curves of SNe Ia are different. The average dispersion on the Hubble Diagram for different samples of SNe Ia is $0.4-0.6^m$. In addition, some discovered SNe Ia are characterized by a red colour at maximum light and low luminosity. The first object with such characteristics was SN 1991bg, which exhibited a rapid decline after maximum light. SN 1991bg was about 2 magnitudes dimmer than other SNe Ia in the Virgo cluster, where the supernova host galaxy, NGC 4374, was located. Objects that are similar to SN 1991bg are called 'Type 1991bg supernovae'.

Some SNe Ia, on the contrary, have high-peak luminosities and slow declines after maximum light. The prototype for this class of supernovae is SN 1991T. The smallest Subclass is the peculiar SNe Iax with absolute magnitudes in the range of $-14.2^m \geq M_{V\,peak} \geq -18.9^m$ but having the same shaped light curves as normal SNe Ia (Foley et al., 2013).

Figure 6 shows the contribution of SN Ia subtypes in volume up to z = 0.08. This choice of volume is motivated by the fact that only for the nearby supernovae are there good spectral data, and selection effects are not significant. However, they cannot be completely avoided. For example, only about thirty SNe Iax were discovered because of their weakness, while it is expected that there are 31 + 17/–13 for every 100 SNe Ia (Foley et al., 2013).

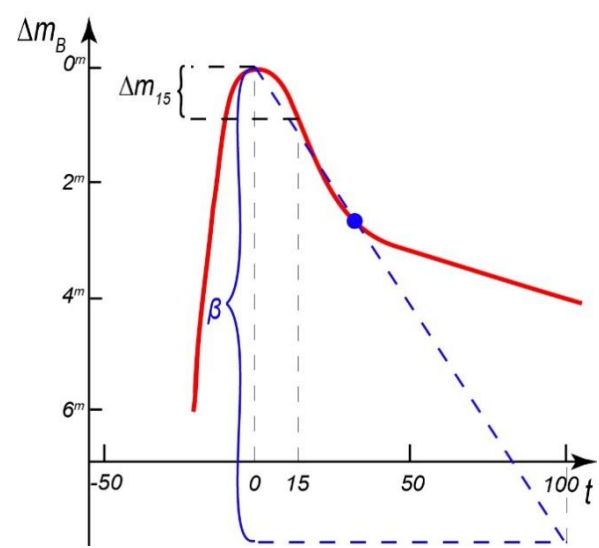

Figure 5: The typical light curve of an SN Ia in the B-filter. The β parameter introduced by Pskovskii and $\Delta m_{15}$ parameter introduced by Phillips are shown. The blue point is the point at which the decline in brightness begins to slow down (Drawn by authors).

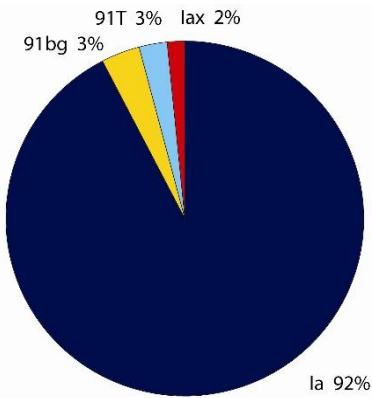

Figure 6: The distribution of SNe Ia subtypes in volume up to z = 0.08. The data are taken from the SAI supernovae catalog (Bartunov et al., 2007) and (Foley et al., 2013) (Drawn by authors).

Thus, mass observations of SNe Ia raised questions about the universality of their light curves, and the 'standard candle' hypothesis was destroyed. But it turned out that there is a relationship between the physical characteristics and the parameters of the light curve of a supernova: the brighter the supernova is, the slower its luminosity declines after the maximum. Therefore, SNe Ia are 'standardizable' objects.

In the 1940s to describe the light curves of novae Dean Benjamin McLaughlin (1901–1965; Lindner, 2014) introduced the $t_3$ parameter—the time in days following the maximum light, during which the nova decreased in brightness by $3^m$—and he found a connection between $t_3$ and the absolute magnitude of novae at maximum ($M_{max}$) (McLaughlin, 1945). To test the idea of Iosif Samuilovich Shklovsky (1916–1985; Gurshtein, 2014) about the absence of a qualitative difference between novae and supernovae, Ivan Mikheevich Kopylov (1928–2000) plotted the $t_3 - M_{max}$ relationship for supernovae (Kopylov, 1955a; 1955b). If Shklovsky had been right, this dependence for supernovae would have been simply a continuation of the dependence for novae. However, the supernovae and novae data points did not follow the same straight line, and the lines associated with each type had different inclinations (see Figure 7). Thus, Kopylov showed that novae and supernovae were two independent types of objects. In this work Kopylov did not divide supernovae by types. However, if one identifies the points on the graph with supernovae whose types were known at that time, an interesting detail is found: Type I SNe (the green line) are positioned differently to all other supernovae. Thus, had Kopylov divided his sample of SNe into Types I and II he would have been the first to find that bright Type I SNe decline more slowly after reaching peak luminosity.

In 1967 Yuri Pavlovich Pskovskii (Figure 8) began to explore similar dependencies for different types of SNe. As a main parameter characterizing the light curve shape, Pskovskii used $β$—the mean rate of decline in photographic brightness from maximum light to the point at which the luminosity decline rate changes. $β$ is measured in magnitudes per 100-day intervals (see Figure 5, and Pskovskii, 1967). The point at which the luminosity decline rate changes, refers to the moment when the rapid luminosity decline is replaced by a slower one. It happens about 25–30 days after the maximum. Selection of this parameter is justified by the fact that, at that time, the probability of discovering a supernova before the maximum light, and obtain the full light curve, was small. Moreover, the existing light curves were mostly incomplete. On the other hand, to determine the decline after the maximum light was rather simple for most observed supernovae. However, first of all, by plotting the $\beta - M_{max}$ relationship Pskovskii, just as Kopylov had done earlier, sought to emphasize the difference between novae and supernovae. Second, upon analyzing Type I SNe, Pskovskii[7] showed that the majority of them has similar values of $\beta$, and therefore their light curves could effectively be considered as 'identical', so Type I SNe could serve as reliable distance indicators.

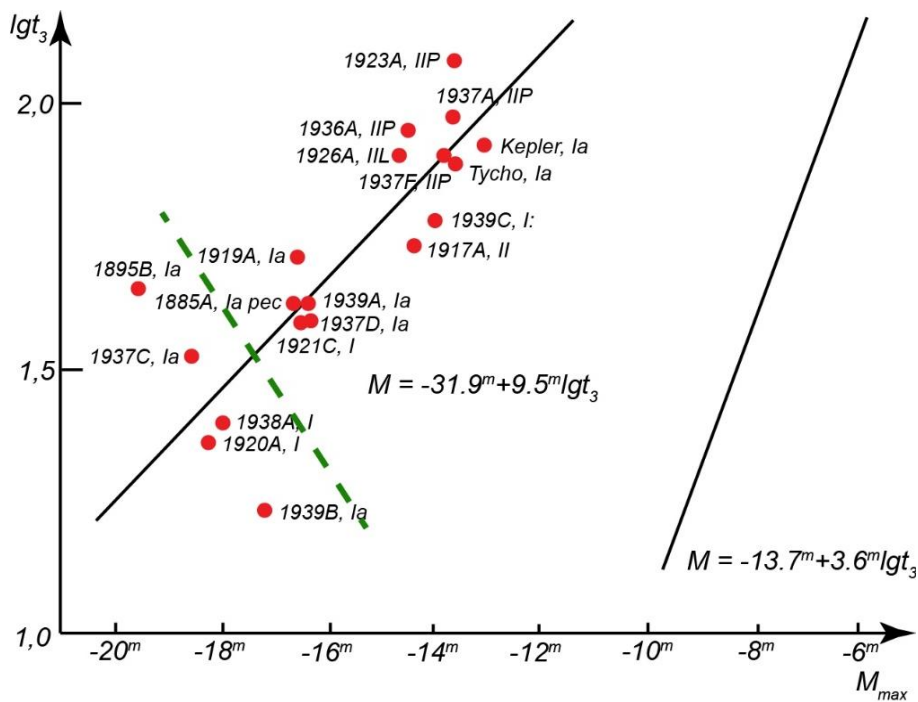

Figure 7: The dependence between the absolute magnitude in maximum light and the decline rate after maximum for SNe (left) and novae (right) (after Kopylov, 1955a; 1955b). The green dashed line shows the plot for SNe Ia. Kepler's and Tycho's historical SNe do not follow the common tendency due to the large uncertainty in their absolute brightness.

In 1973 Roberto Barbon and his collaborators subdivided Type I SNe into two classes, depending upon their decline rates: 'fast' and 'slow' (Barbon et al., 1973). 'Fast' Type I SNe were brighter at maximum than the 'slow' ones. In addition, Barbon et al. concluded that the existence of two subclasses of Type I SNe is physically justified because there was a connection between the SN subclass and the type of host galaxy: 'fast' Type I SNe avoided elliptical galaxies while 'slow' Type I SNe avoided irregular galaxies. Further, Barbon et al.'s studies (1975) showed that there was no significant difference between 'fast' and 'slow' Type I SNe.

In 1977 Pskovskii published a paper in which he proposed the introduction of a photometric classification of SNe based on the value of $\beta$: SNe with large values of $\beta$ would be called 'senior' and those with small values of $\beta$ 'junior' (Pskovskii, 1977). Photometric classes were denoted as follows: SN Type followed by the value of $\beta$ after the decimal point; for example, photometric class I.10 means that the SN belongs to Type I and as $\beta$ value of 10. In the same paper Pskovskii showed that there was a relationship linking the absolute magnitude at maximum for Type I SN with the $\beta$ parameter. On the basis of 32 Type I SNe this relation was found to be

$$-21.3 + 0.11\beta = M_{pg} \pm 0.5 \qquad (3)$$

where $M_{pg}$ is the photographic magnitude (Pskovskii, 1977). Thus, Pskovskii came to the correct conclusion that supernovae with a slow decline rate were brighter than supernovae with a fast decline rate. Later, he confirmed this relation using an expanded sample of Type I SNe (Pskovskii, 1984).

However, it should be noted that in 1974 the American statistician and astronomer Bert Woodard Rust (b. 1940; Figure 9) independently derived the correct relation between the peak luminosity and the light curve slope for Type I SNe in his Ph.D. thesis *The Use of Supernovae Light Curves for Testing the Expansion Hypothesis and Other Cosmological Relations.* To characterize the slope of the light curve Rust used the parameter

$$\Delta t_c = t(m_0 + 0.5) - t(m_0 + 2.5) \qquad (4)$$

where $m_0$ is the supernova magnitude at maximum. To determine the absolute magnitude of SNe Rust (1974) used the following formula:

$$M_0 = (-18.5 \pm 0.68) - (0.0512 \pm 0.0359)\Delta t_c \qquad (5)$$

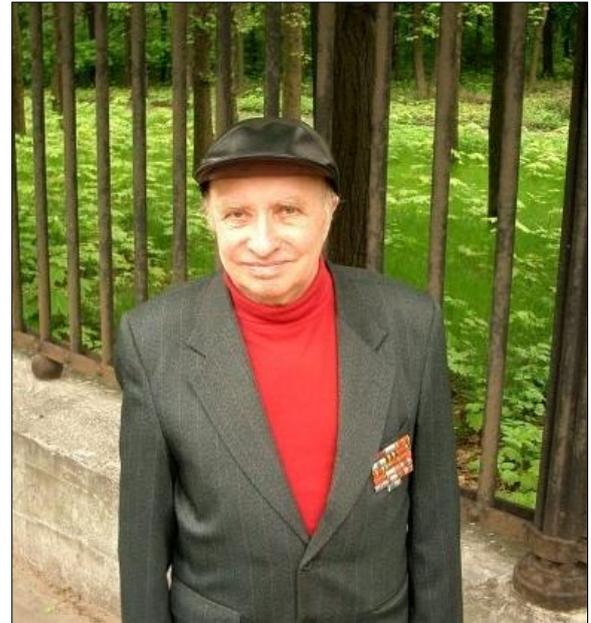

Figure 8: Yury Pavlovich Pskovskii, 1 February 1926–21 July 2004 (http://www.astronet.ru/db/msg/1211317).

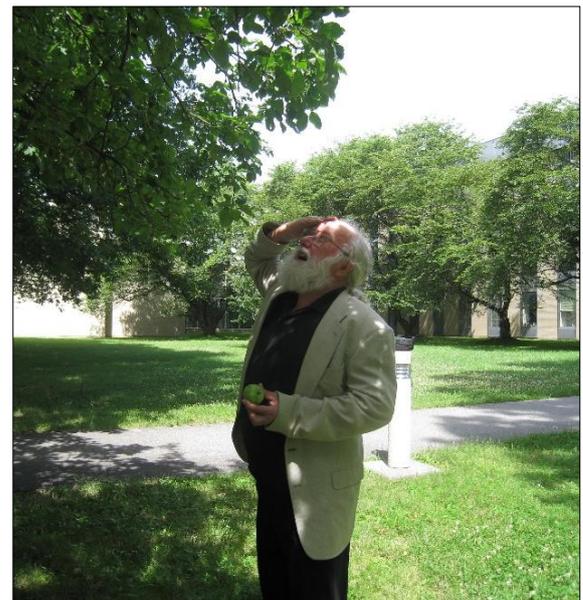

Figure 9: Professor Bert Woodard Rust, who works at the National Institute of Standards and Technology, in the USA (Photograph courtesy Professor Rust from his personal archive).

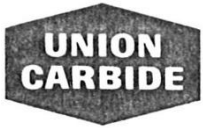

UNION CARBIDE CORPORATION

NUCLEAR DIVISION

P.O. BOX X, OAK RIDGE, TENNESSEE 37830

October 24, 1973

Dr. Yu. P. Pskovskij
Sternberg State Astronomical Institute
University of Moscow
Universitetskij Prospect 13
Moscow V-234, 117234
USSR

Dear Dr. Pskovskij:

    This letter is to thank you for your interest in my work. I am especially pleased because your own work has been so important for the success of mine. I am referring to your three papers, "The Photometric Properties of Supernovae," "Phase Dependence of the Colors of Type I and II Supernovae," and "Photometric Aspects of Type I Supernovae." Since I can not read Russian, I was forced to use the translations which appeared in the translation journal Soviet Astronomy - AJ. But even in translation, the importance of the work is very apparent, and I must take this opportunity to express my admiration for it.

    My own work is concerned mainly with cosmological applications of supernovae light curves. A problem that I have encountered many times is that of estimating the maximum luminosity and time of maximum using a fragmentary light curve. Another problem is that of correcting the magnitudes for absorption. Your work on light curves and color curves provided just the tools I needed!

    I have not yet quite completed my dissertation which gives an account of all this. I have written seven chapters and must write two more in order to complete it. The writing is proceeding rather slowly because I am employed full-time as a mathematician at Oak Ridge National Laboratory. I hope to have it completed by June 1974. I will send you a copy as soon as it is finished. I will also send you a reprint of the PASP article as soon as I receive them. If you have any other papers of your own on the subject of supernovae, I would be very pleased to receive reprints of them.

    Thanking you again for your interest in my work, I am

                                 Yours sincerely,

                                 Bert W. Rust
                                 Computer Sciences Division

BWR/bm

Figure 10: A 1973 letter from Bert Rust to Yuri Pskovskii, 1973 (courtesy Professor Rust from his personal archive).

Never having met in person, Rust and Pskovskii were in long-term scientific and friendly correspondence. In one of the letters to Pskovskii, Rust wrote that he was interested in the problem of determining the luminosity of supernovae at maximum light by using a fragmentary light curve (see Figure 10). Undoubtedly, Bert Rust was the first to discover this important relationship. However, as Rust noted in one of his letters, Pskovskii's work had a tremendous impact on his own research.

Unfortunately, at the time Rust's discovery went unnoticed by the astronomical community. This was partly due to the fact that Rust's publications on this topic were not in high-profile astronomical journals. Except in his Ph.D. thesis, Rust (1974) mentioned this correlation only once, in a conference abstract that was published in the *Bulletin of the American Astronomical Society* (Rust, 1975).

In 1981 David Branch studied the possibility of sub-classifying Type I SNe by photometric properties and showed that the light curve parameters of these SNe are distributed in a continuous manner, and do not form two subclasses as Barbon had claimed. In addition, Branch (1982) confirmed Pskovskii's conclusion that the absolute magnitude at maximum light is proportional to the decline rate after the maximum. However, further analyses of SNe Ia, conducted by Douglas L. Miller and David Branch (1990), did not reveal this relationship.

John R. Boisseau and J. Craig Wheeler (1991) then explored the question of how the background light from the host galaxies of SNe Ia may affect the observed changes in absolute magnitude at maximum light, and the decline rate. By adding a small amount of the background of the host galaxy to the photometric data, they noticed both an increase in the peak luminosity and a flattening of the light curve, in other words, the Rust-Pskovskii relationship. They showed that the contribution from the background becomes more significant for the faint objects. Thus, the correct background accounting is particularly important for the study of distant SNe Ia, because their light contains a large degree of contamination by the background light of the host galaxy. Boisseau and Wheeler (ibid.) came to the conclusion that the observed dispersion of parameter $\beta$ is random, and that most SNe Ia have similar light curves.

As can be seen, the Rust-Pskovskii relationship repeatedly was subjected to inspection and criticism. But nowadays it is the most important relationship in observational cosmology that is based on the study of distant SNe Ia.

In the early 1980s CCD cameras appeared, and the number of SNe discoveries increased substantially. Moreover, the probability of discovering SNe before they reached maximum light and following their brightness evolution longer also increased. The first light curves of SNe Ia obtained using CCD photometry showed that some supernovae had faster decline rates than others. Later, the low luminosity SN Ia 1991bg with a fast decline rate was discovered. All this motivated the American astronomer Mark Phillips to revise the Rust-Pskovskii relationship using nine SNe Ia with well-known distances measured either using the Tully-Fisher Relation or the surface brightness fluctuation method for galaxies. Since the point where the luminosity decline rate changes (and therefore, the $\beta$ parameter) is difficult to determine with high accuracy, as an alternative to the $\beta$ parameter Phillips used $\Delta m_{15}$—a parameter that indicates how many magnitudes fainter the luminosity becomes in blue light during the first 15 days after maximum light (see Figure 5). Parameter $\Delta m_{15}$ initially was proposed by George Jacoby, as Phillips (1993) noted in the acknowledgements in his paper. The relation between the absolute magnitude at maximum light in the *B*-band and $\Delta m_{15}$, derived by Phillips (ibid.) was

$M_{Bmax} = -21.726(0.498) + 2.698(0.359)\, \Delta m_{15(B)}$   (6)

The use of $\Delta m_{15}$ reduced the dispersion of $M_{Bmax}$ by a factor of two. There is also a quadratic relation between the absolute magnitude at maximum light and the slope parameter $\Delta m_{15}$.

## 4 DISCUSSION

The existence of empirical relationships between the luminosity and light curve shape of SNe Ia is explained in some theoretical models (e.g. see Höflich and Khokhlov, 1996; Höflich et al., 1993; Livne and Arnett, 1995; Woosley and Weaver, 1994). It is now generally accepted that the SN Ia phenomenon arises from an explosion of a carbon-oxygen white dwarf with a mass close to the Chandrasekhar Limit. Different theoretical models include deflagration (subsonic combustion), detonation (supersonic combustion), delayed detonation, off-center detonation, pulsating delayed detonation etc. (Arnett, 1969; Khokhlov, 1991; Nomoto et al., 1976; Ruiz-Lapuente et al., 1993). There is also a model with a sub-Chandrasekhar mass white dwarf as a progenitor of an SN Ia, wherein an explosion occurs on the surface of the white dwarf due to the ignition of the helium layer accumulated as a result of the accretion. One possible explanation of the observed relationship is that the density at which the detonation combustion is replaced by deflagration affects the amount of $^{56}$Ni (the decay of which is responsible for the light curve shape of SNe Ia) synthesized in the explosion (Blinnikov and Tsvetkov, 2009). If the change of combustion modes occurs relatively late, then the outer envelope is able to expand, reducing the amount of produced $^{56}$Ni. This leads to a decrease in temperature of the expanding envelope and photospheric opacity decreases rapidly. Therefore, the photosphere becomes transparent earlier and the energy is released in a short period of time. Conversely, if the detonation replaces deflagration early enough, a large amount of $^{56}$Ni is produced. The result is a bright hot supernova whose opaque envelope loses energy rather slowly, which explains the slow decline rate in the luminosity for the brightest supernova light curves.

Theoretical models of the SN Ia explosion only partly explain the heterogeneity of supernovae and the origin of the relationship between the peak luminosity of SNe Ia and their luminosity decline rate after maximum light. It remains to be seen whether all or some of the proposed theoretical models can be explained by the variations exhibited.

## 5 CONCLUDING REMARKS

It is generally believed that SNe Ia are good 'standard candles' and can serve as distance indicators in the Universe. However, their peak luminosity differs from one SN to another. The situation was saved by the discovery of a relationship between the luminosity of SNe Ia and their decline rate after the peak brightness– a slower decline corresponds to a brighter SN. This idea originally was proposed by Bert Rust and Yuri Pskovkii in the 1970s. However, at that time the number of well-studied SNe was small and their light curves were essentially incomplete. Probably this fact was the reason why Pskovskii used the $\beta$ parameter to characterize the light curves of SNe. For the observed SNe, at that time it was easier to measure the initial decline in the light curve rather than catch a SN before it reached maximum light and plot the entire light curve.

In 1993 Mark Phillips used better observational data to revise the idea proposed by Rust and Pskovskii. In his method he successfully linked the absolute magnitude of SNe Ia with a parameter $\Delta m_{15}$ which indicates how many magnitudes fainter luminosity becomes in blue light during the first 15 days after the maximum light. Phillips' study revealed the same trend as previous ones.

From this time, several different concepts of the 'standardization' of SNe Ia were developed: $\Delta m_{15}$ (Phillips, 1993; Phillips et al., 1999); stretch-factor (Perlmutter et al., 1997; 1999); Multicolor Light Curve Shape (Jha et al., 2007; Riess et al., 1996;); PRES (Prieto et al., 2006); Spectral Adaptive Light curve Template for Type Ia supernova (Guy et al., 2005; 2007); Color-Magnitude Intercept Calibration (Wang et al., 2003), etc. All of these 'standardization' methods allowed the determination of the distance to SNe Ia based on the relationship between various parameters, depending on the distance (maximum brightness or the average difference in magnitudes between the observed light curve and the reference light curve) or parameters that did not depend on the distance (the Colour Index, stretch-factor or $\Delta m_{15}$). Applying standardization techniques for SN Ia light curve analysis produced an improved value for the cosmological constant.

The development of new, more accurate, standardization techniques in order to use SNe Ia as reliable distance indicators is still an important task, and one which the international astronomical community is addressing in current research.

## 6 NOTES

1. Today S Andromedae and Z Centauri are classified as supernovae SN 1885A and SN 1895B respectively.
2. This, however, was not a cornerstone in Shapley's theory. Unlike Curtis, he was correct in placing the Solar System far out from the center of our galaxy and in his suggestion that the size of our galaxy was much larger than previously estimated. Shapley's and Curtis' estimations, of $\geq 3 \times 10^5$ and $\leq 3 \times 10^4$ l.y. respectively, were almost equally erroneous in comparison to the modern value of $10^5$ l.y.
3. Although Osterbrock (2001) states that this term was first used in print in 1933 in a publication by the Swedish astronomer Knut Emil Lundmark (1889–1958; Teerikorpi, 2014), according to Zwicky (1940) it had been used from 1931 in seminars and courses given by Zwicky and Baade at Caltech.
4. The average absolute magnitude is many times fainter than the currently-accepted value because Baade used $H_0$ = 500 km/s/Mpc.
5. Rudolph was a nephew of a famous mathematician Hermann Minkowski (1864–1909; Hall, 2014).
6. Type I SNe spectra were decrypted later by Soviet astronomer Yu.P. Pskovskii in 1969, however the first astronomer who tried to explain the minima in Type I SNe spectra as absorption lines was Dean McLaughlin, in 1963.
7. Subtypes Ia, Ib and Ic were identified later.

## 7 ACKNOWLEDGEMENTS

The authors thank Professor Bert Woodard Rust for providing material from his personal archives and especially for his useful comments and in-teresting discussion about cosmology. Maria Pruzhinskaya acknowledges Professors V.M. Lipunov and S.I. Blinnikov for their encouragement and D.Yu. Tsvetkov for his comments that helped improve the manuscript. We acknowledge Katharine Mullen for her support, comments and writing assistance. We also thank the referees for their corrections and suggestions and Professor P. Lundqvist.

M.V. Pruzhinskaya acknowledges the support of the Mechnikov Scholarship from the French Government for a 3-month research visit to the Observatoire de la Côte d'Azur where part of this work was completed. Sergey Lisakov is supported by the Erasmus Mundus Joint Doctorate Program by Grants Number 2013-1471 from the agency EACEA of the European Commission.

Maria Victorovna Pruzhinskaya (below, left) was born in Syktyvkar (Russia) in 1988. She graduated from Lomonosov Moscow State University with Honours in 2011 and received a Ph.D. from this University in 2014. As a Ph.D. student she has been working in the MASTER Global Robotic Net project (Mobile Astronomical System of TElescope-Robots, http://observ.pereplet.ru/). Her Ph.D. project "Supernovae, gamma-ray bursts and accelerating expansion of the Universe" and recent research cover several aspects of supernovae and gamma-ray burst theory and observations. Maria works at the Sternberg Astronomical Institute of Lomonosov Moscow State University and at the Laboratoire de Physique Corpusculaire de Clermont-Ferrand in France.

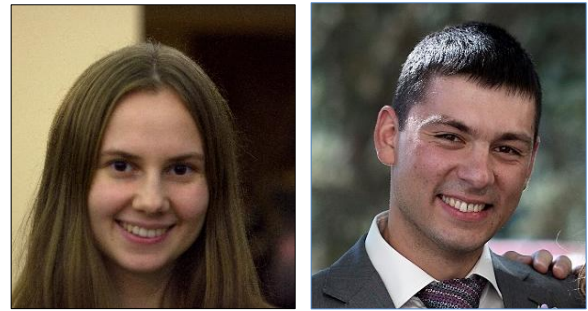

Sergey Mikhaylovich Lisakov (above, right) was born in Leninsk, Kazakh SSR in 1990. His father was a military specialist working at Baikonur Cosmodrome. In 1992 the family moved to Moscow (Russia), where Sergey graduated from the Astronomy Department in the Faculty of Physics at Lomonosov Moscow State University. In September 2013 he started his Ph.D. research on "Core-collapse supernovae and their progenitors" at the University of Nice Sophia Antipolis in France.